\documentclass[b5paper,12pt]{iopart}
\usepackage{graphicx}

\begin{document}

\title[Baryon-Strangeness correlations in Parton/Hadron transport model]{Baryon-Strangeness correlations in
Parton/Hadron transport model for Au + Au collisions at $\sqrt{s_{NN}}$ = 200 GeV}
\author{F. Jin$^{1,2}$, Y. G. Ma$^{1}$\footnote[7]{Corresponding author. Email:
ygma@sinap.ac.cn}, G. L. Ma$^1$, J. H. Chen$^{1,2}$, S.
Zhang$^{1,2}$, X. Z. Cai$^1$, H. Z. Huang$^3$,
J. Tian$^{1,2}$, C. Zhong$^1$, J. X. Zuo$^{1,2}$}
\address{$^1$ Shanghai Institute of Applied Physics, Chinese Academy of Sciences,
P.O.Box 800-204, Shanghai 201800, China}
\address{$^2$ Graduate School of the Chinese Academy of Sciences,
Beijing 100080, China}
\address{$^3$ Deptartment of Physics and Astronomy, University of California, Los Angeles,
CA 90095, USA}

\begin{abstract}

Baryon-strangeness correlation (C$_{BS}$) has been investigated
with a multi-phase transport model (AMPT) in $^{197}$Au +
$^{197}$Au collisions at $\sqrt{s_{NN}}$ = 200 GeV. The centrality
dependence of C$_{BS}$ is presented within the model, from
partonic phase to hadronic matter. We find that the system still
reserve partial predicted signatures of C$_{BS}$ after parton
coalescence. But after hadronic rescattering, the predicted
signatures will be obliterated completely. So it seems that both
coalescence hadronization process and hadronic rescattering are
responsible for the disappearance of the C$_{BS}$ signatures.

\end{abstract}

\submitto{\JPG}

\section{Introduction}\label{intro}

Ultra-relativistic heavy ion collision may provide sufficient
conditions for the formation of a deconfined plasma of quarks and
gluons~\cite{Karsch_NPA698}. Experimental results from RHIC
indicate that a strongly-interacting partonic matter (termed sQGP)
has been created in the early stage of central Au + Au collisions
at $\sqrt{s_{NN}}$ = 200 GeV at RHIC~\cite{BRAHMS_NPA757}. In
order to uncover the nature of this matter, probes based on
fluctuations have been proposed throughout the last
decade~\cite{Stodolsky_PRL75,Shuryak_PLB423,Bleicher_NPA638}.
Recently a novel event-by-event observable has been introduced by
Koch et al.~\cite{Koch_PRL95}, i.e. the baryon-strangeness
correlation coefficient C$_{BS}$.

The correlation coefficient C$_{BS}$ is defined as

\begin{equation}
C_{BS} \; = \;
-3\frac{ {\sigma}_{BS}  }{{\sigma}_S^2  }
\; = \;
-3\frac{ \langle BS \rangle - \langle B \rangle \langle S \rangle }{ \langle S^2 \rangle - {\langle S \rangle}^2 }
\label{eq:bs}
\end{equation}
where $B$ is the baryon charge and $S$ is the strangeness in a
given event.

This correlation is proposed as a tool to specify the nature of
the highly compressed and heated matter created in heavy-ion
collisions~\cite{Haussler_PRC73}. The idea is from that the
relation between baryon number and strangeness will be different
when the phase of system is different. On the one hand, if the
basic degrees of freedom are weakly interacting quarks and gluons,
the strangeness is carried exclusively by the $s$ and \=s quarks,
B$_{S}$=-$\frac{1}{3}$S$_{S}$. Thus the correlation coefficient
C$_{BS}$=1. On the other hand, if the degrees of freedom are
hadronic matter, the case is different because the
baryon-strangeness correlation coefficient strictly depends on the
hadronic environments. For example, in a system composed of kaons
the coefficient C$_{BS}$ $\approx$ 0, but C$_{BS}$ $\approx$ 1.5 for
Cascades system.

In this article, we study the correlation coefficient with the
AMPT model~\cite{BZhang_PRC72}, which
consists of four main components: the initial conditions, partonic
interactions, hadronization and hadronic rescattering. The initial
conditions, which include the spatial and momentum distributions
of minijet partons and soft string excitations, are obtained from
HIJING model~\cite{XNWang_PRD44}. In the default version of AMPT
model (i.e. the default AMPT)~\cite{BZhang_PRC61}, minijet partons
are recombined with their parent strings when they stop
interactions. Then the resulting strings and the initial excited
strings are converted to hadrons using the Lund string
fragmentation model~\cite{BAnderson_PR97}. In the string melting
version of the AMPT model (i.e. the string melting
AMPT)~\cite{ZWLin_PRC65}, the initial matter fragments into
partons. A quark coalescence model is used to combine partons to
form hadrons. In the two versions, scatterings among partons
including the resulting partons and the initial minijet partons
are modelled by Zhang's parton cascade model
(ZPC)~\cite{BZhang_CPC109}, meanwhile, dynamics of the
hadronic matter is described by A Relativistic Transport (ART)
model~\cite{BALi_PRC52}. Details of the AMPT model can be found in
a recent review~\cite{BZhang_PRC72}. Previous
studies~\cite{GLMa_PLB647,JHChen_PRC74} demonstrated that the
partonic effect cannot be neglected and the string melting AMPT
model is much more appropriate than the default AMPT model in
describing nucleus-nucleus collisions at RHIC. In the present
work, the parton interaction cross section in the AMPT model is
assumed to be 10 mb which is the same as we used in our previous
publications ~\cite{GLMa_PLB647,JHChen_PRC74,SZhang_PRC76}.

\section{Results}\label{result}

Because the default AMPT is based on string mechanisms it provides
an estimate of C$_{BS}$ value in the case where no partonic matter
is created. And the string melting AMPT is based on strong parton
cascade, therefore it provides an estimate of C$_{BS}$ value as
the partonic matter is created. So we can compare the values of
C$_{BS}$ in the two model versions  to learn information about
partonic matter at RHIC.

Firstly, we study the partonic phase with the string melting AMPT.
we will choose appropriate pseudorapidity windows and study the effect
of parton cascade.

\begin{equation}
C_{BS} \; = \;
-3\frac{ {\sigma}_{BS}  }{{\sigma}_S^2  }
\; = \;
1+\frac{ {\sigma}_{us}^2  }{ {\sigma}_S^2  }+ \frac{ {\sigma}_{ds}^2  }{ {\sigma}_S^2  }
\label{eq:bs2}
\end{equation}

In an uncorrelated partonic phase, ${\sigma}_{us}^2$ $\approx$ 0
and ${\sigma}_{ds}^2$ $\approx$ 0, we get $C_{BS}$ $\approx$ 1.
The results from the lattice QCD also predicted $C_{BS}$ $\approx$
1 above $T_{C}$. So models of the deconfined matter should obey
such constraints.

\begin{figure}[htb]
\vspace{-2pt}

\resizebox{0.5\textwidth}{!}{\includegraphics{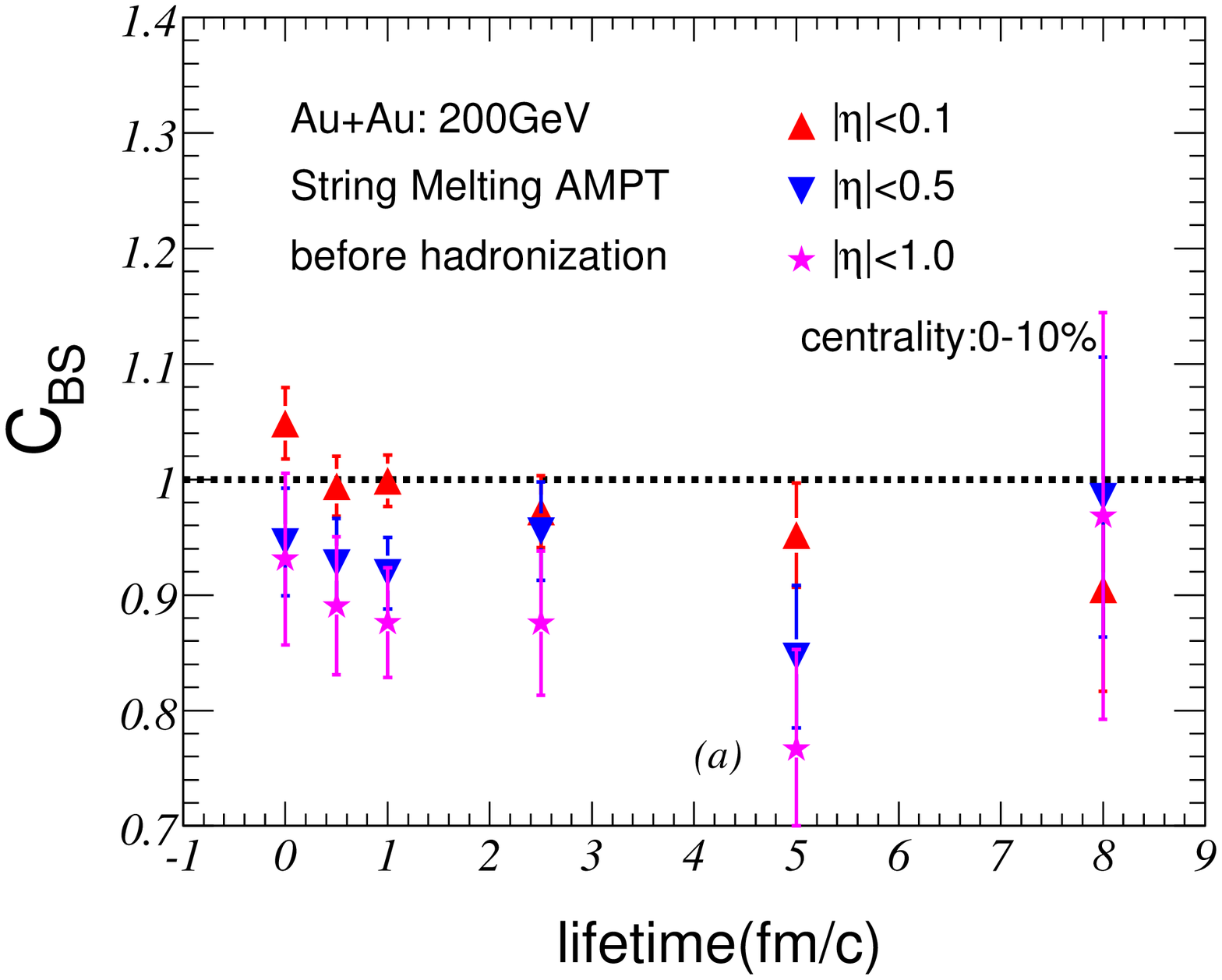}}
\resizebox{0.5\textwidth}{!}{\includegraphics{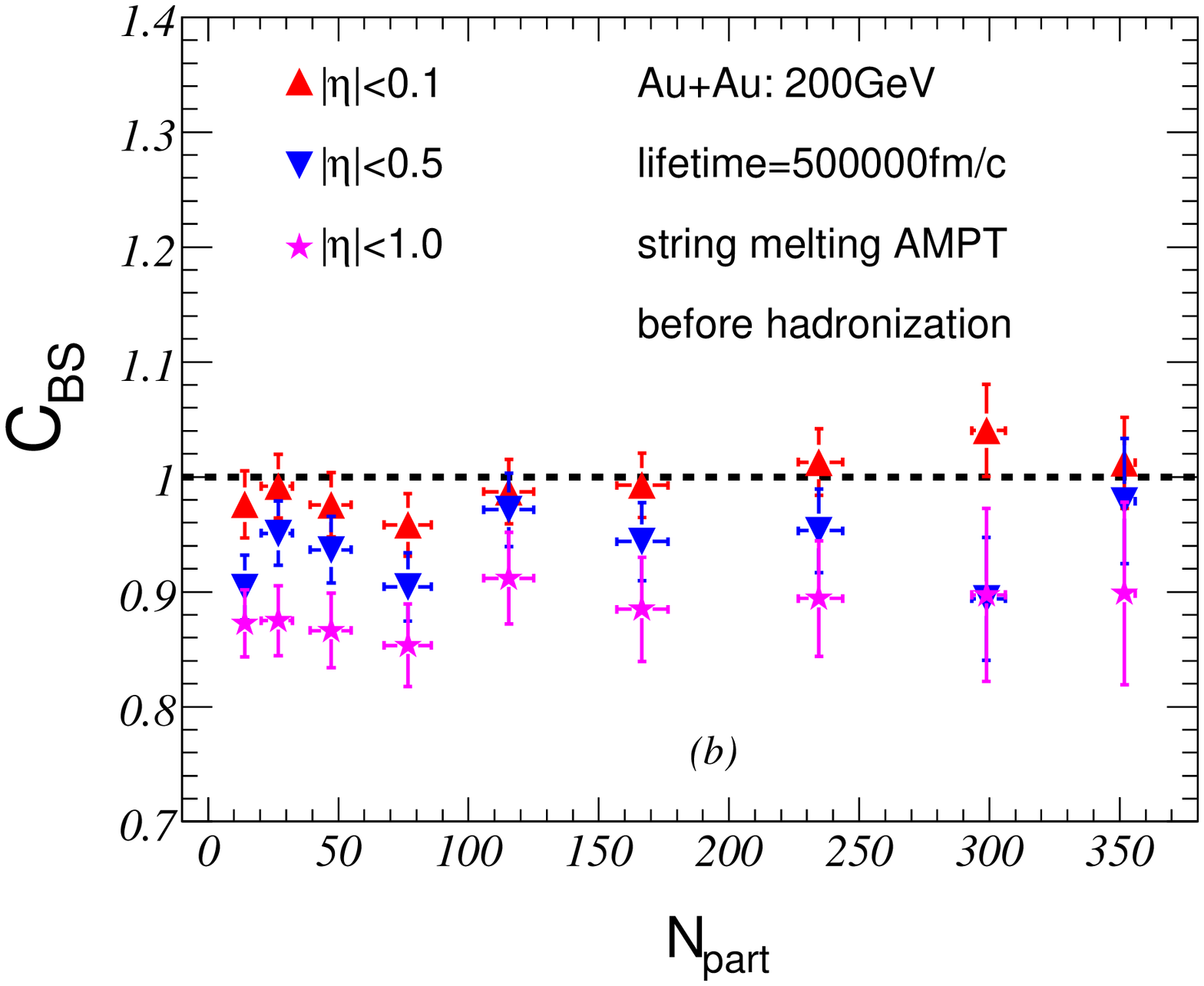}}

\caption[]{(a) The time evolution of baryon-strangeness
correlation coefficient $C_{BS}$ for partonic matter at
$\eta_{max}$=0.1, 0.5 and 1.0; (b) The dependence of $C_{BS}$ on
the number of participant particles in different pseudorapidity
windows, namely $\eta_{max}$ = 0.1, 0.5 and 1.0 for an infinite
lifetime of partonic matter.
} \label{fig1}
\end{figure}

Fig.~\ref{fig1} (a) depicts the time evolution of $C_{BS}$ of
partonic matter. We find $C_{BS}$ $\approx$ 1 with increasing time
of parton cascade in different pseudorapidity windows 0.1 and 0.5,
even for an infinite partonic lifetime. Therefore we conclude that
in the above pseudorapidity windows parton cascade does not
influence $C_{BS}$. But Fig.~\ref{fig1} shows that when
$\eta_{max}$ = 1.0, the conditions that ${\sigma}_{us}^2$=0 and
${\sigma}_{ds}^2$=0 are not perfectly satisfied after long parton
cascade period. Therefore, we will focus on the correlations only
in the two pseudo-rapidity windows, namely  $\eta_{max}$ = 0.1 or
0.5. In the following work, we present hadronic $C_{BS}$ including
all hadrons with masses up to that of $\Omega^{-}$.

In AMPT model, hadronization is described with a coalescence
model, and the pseudo-rapidity distribution will change during
this process. After hadronization, the strangeness will be
enhanced, so the $C_{BS}$ will drop. In the following work, we
can see that.

\begin{figure}[htb]
\vspace{-3pt}
\resizebox{0.5\textwidth}{!}{\includegraphics{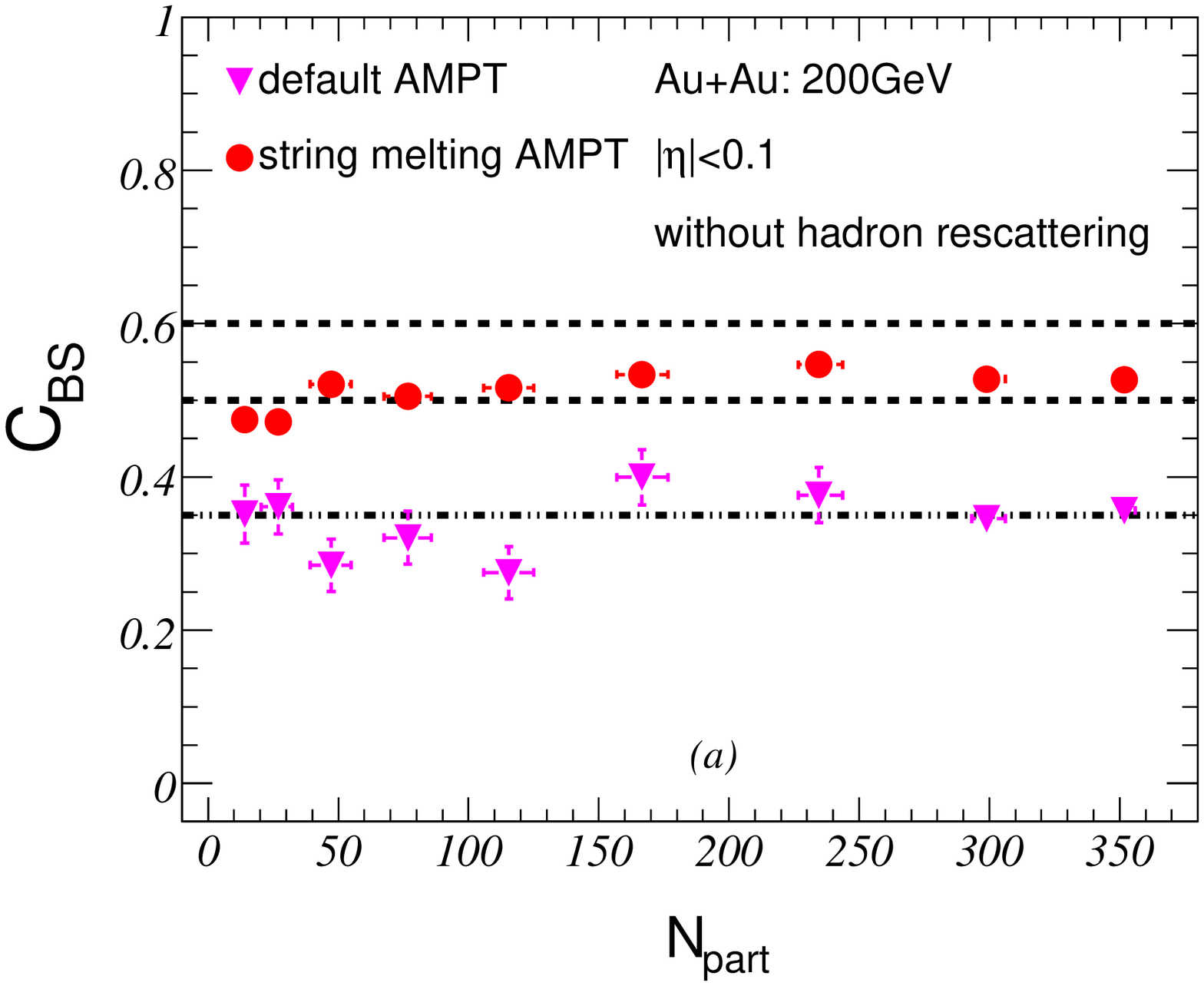}}
\resizebox{0.5\textwidth}{!}{\includegraphics{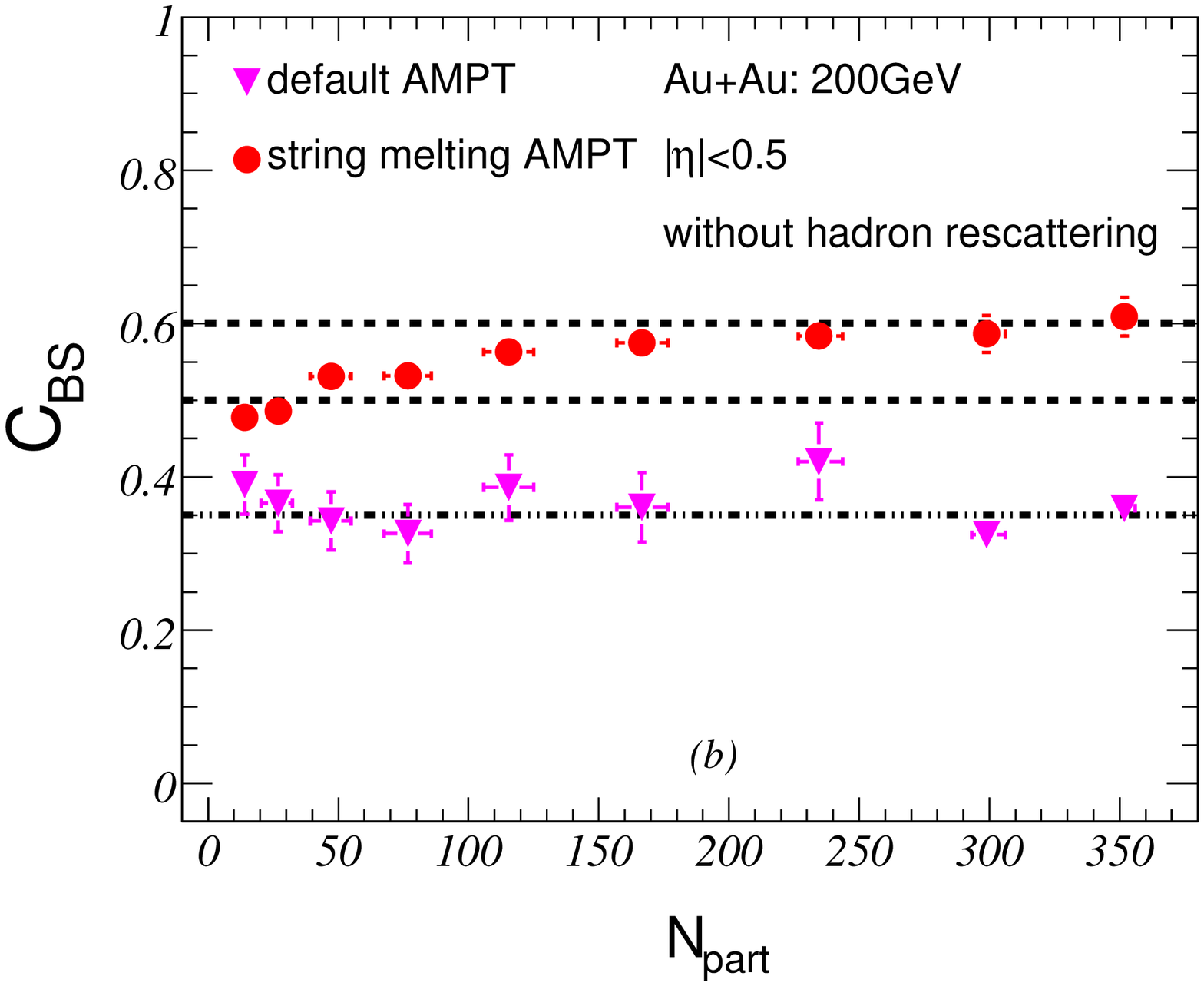}}

\caption[]{ $C_{BS}$  as a function of $N_{part}$ at the
$\eta_{max}$=0.1 (a) and $\eta_{max}$=0.5 (b) in the default AMPT and
the string melting AMPT without hadronic rescattering, respectively. } \label{fig2}
\end{figure}

In Fig.~\ref{fig2}(a) $C_{BS}$ is depicted as a function of the
number of participant particles ($N_{part}$). For small acceptance
windows around mid-pseudo-rapidity, $C_{BS}$ stays roughly
constant. The value of $C_{BS}$ may be estimated as simply the
ratio of probability to observe a strange baryon to that of
strange meson~\cite{Majumder_0}; 0.5 for the string melting AMPT
and 0.35 for the default AMPT. In this case, it is concluded that
if the deconfined phase exists, the ratio of multiplicity of
strange baryon to that of strange meson will be enhanced before
hadron rescattering. For $\eta_{max}$=0.1, the pseudo-rapidity
windows may be too small to include essential correlation
information. So in the following study, we will fix the
pseudo-rapidity window at 0.5. Fig.~\ref{fig2}(b) shows that for a
larger acceptance window $\eta_{max}$=0.5, $C_{BS}$ increases from
peripheral toward central collisions for the string melting AMPT
because of higher baryon density. For central collisions $C_{BS}$
goes up to 0.6 and becomes flat. But for the default AMPT,
$C_{BS}$ still stays roughly constant because of fragmentation
mechanism. The results are consistent with ~\cite{Haussler_PRC73}.
From Fig.~\ref{fig2}(b) we can say there exists an enhanced
$C_{BS}$ for central collisions if there is a partonic phase
before the hadronic rescattering.

\begin{figure}[htb]
\vspace{-3pt}
\resizebox{0.5\textwidth}{!}{\includegraphics{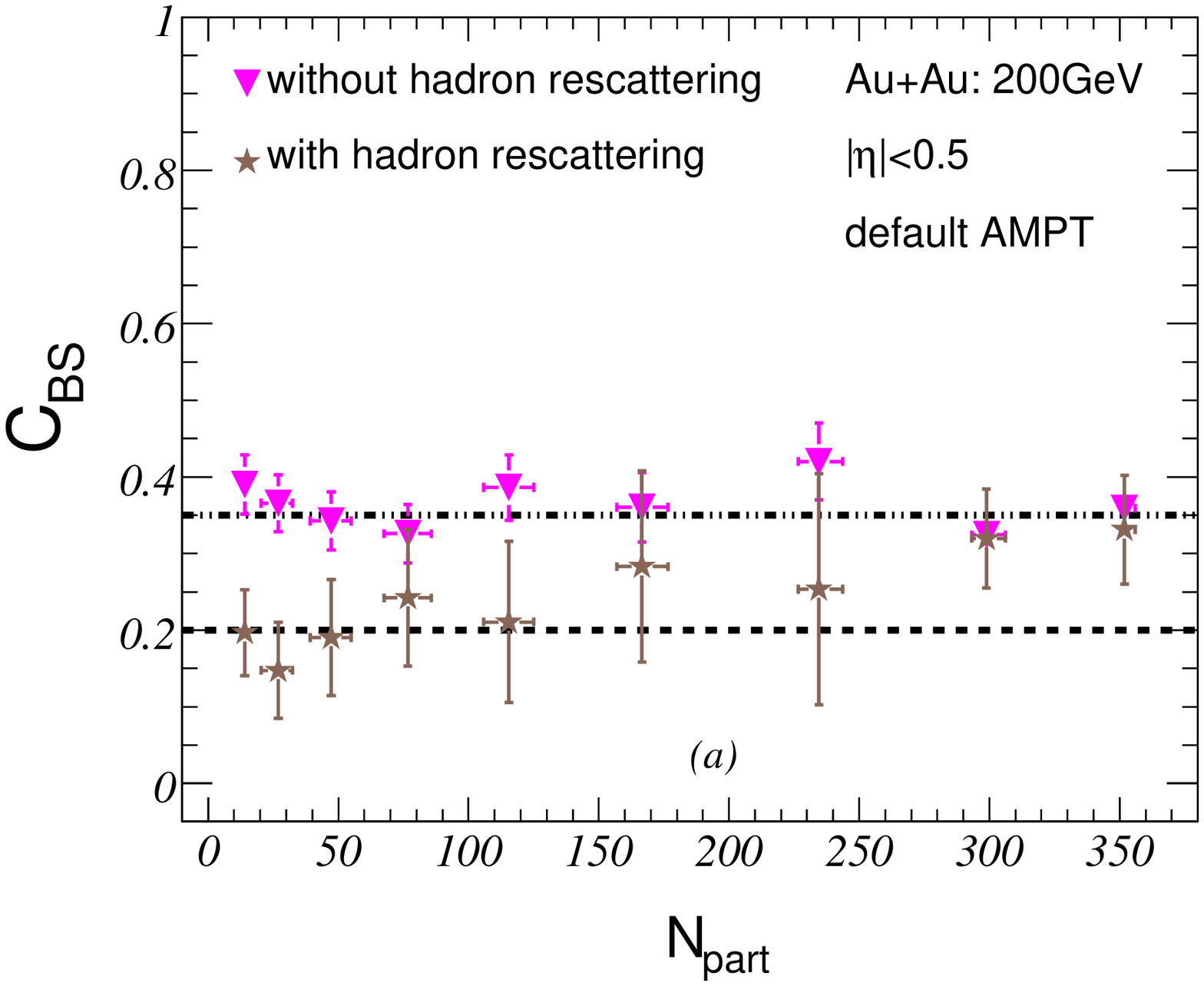}}
\resizebox{0.5\textwidth}{!}{\includegraphics{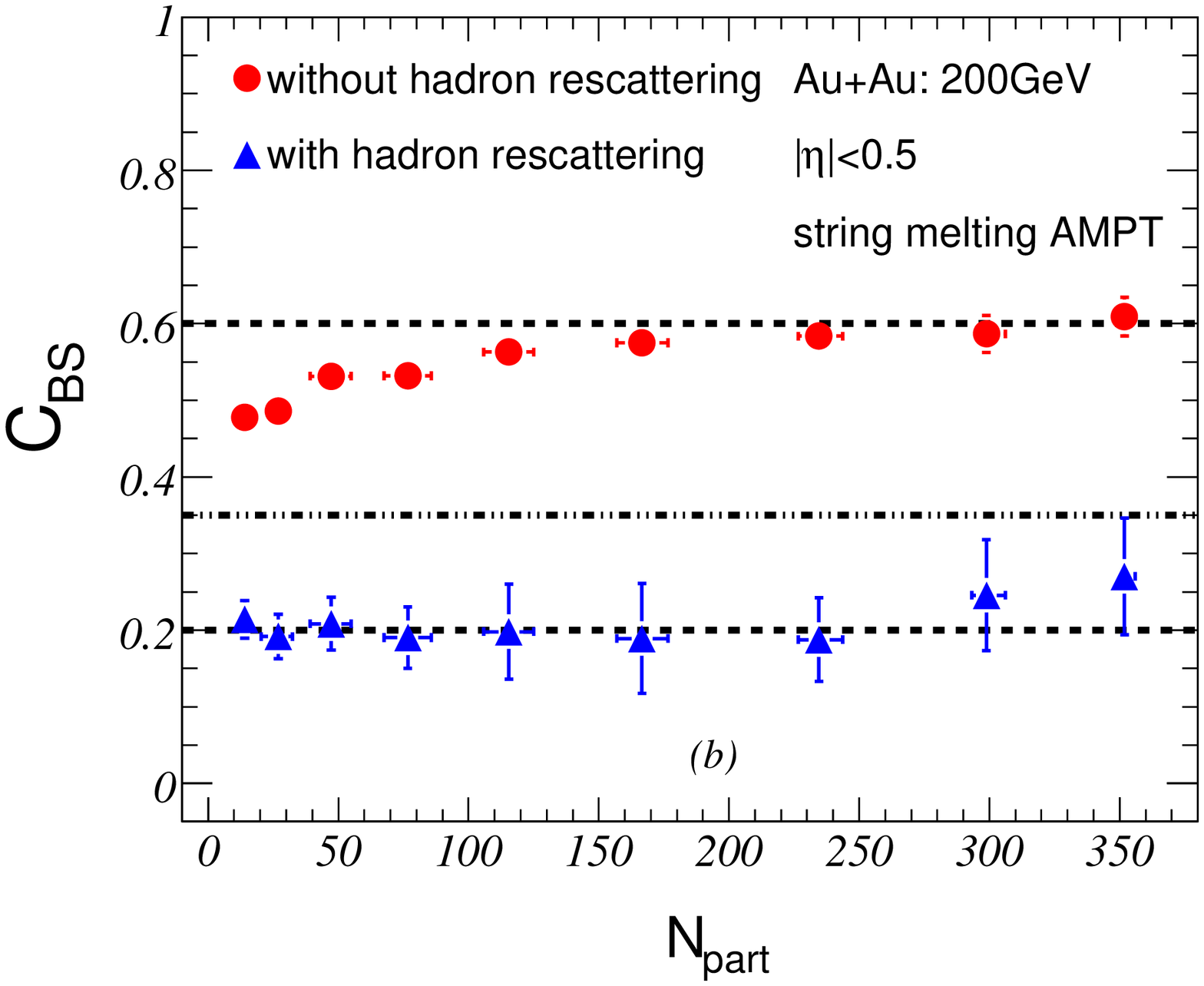}}
\caption[]{ $C_{BS}$ as a function of $N_{part}$  at the
$\eta_{max}$=0.5 in the default AMPT (a) and the string melting AMPT
(b) without hadronic rescattering and with hadronic rescattering. } \label{fig3}
\end{figure}

\begin{figure}[htb]
\centering\mbox{ \vspace*{-3pt}
\includegraphics[width=0.60\textwidth]{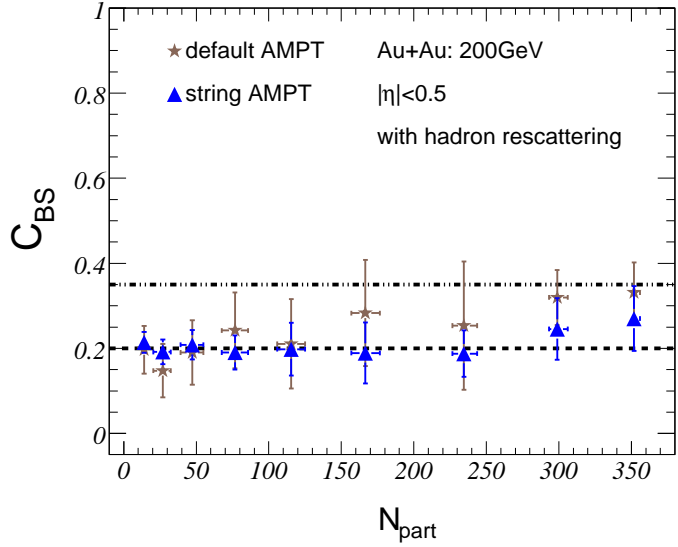}}
\vspace*{-3pt} \caption[]{$C_{BS}$ as a function of $N_{part}$ at
$\eta_{max}$ = 0.5 in the default AMPT and the string melting AMPT
with hadronic rescattering.
 } \label{fig4}
\end{figure}

We also investigate the effect of hadronic rescattering which is
shown in Fig.~\ref{fig3}. We find the hadronic rescattering has a
larger effect on the $C_{BS}$ for the string melting AMPT
simulation than that of the default AMPT from Fig.~\ref{fig3}(b).
For the default case, the hadronic effect on $C_{BS}$ is trivial
as shown in Fig.~\ref{fig3}(a), this is consistent with the
results from UrQMD \cite{Haussler_PRC73}. In central collisions,
$C_{BS} \sim 0.4$ which is also close to UrQMD result. For the
melting case, the hadronic rescattering causes the disappearence
of the signal of partonic matter as shown in Fig.~\ref{fig4}.

\begin{figure}[htb]
\vspace{-3pt}
\resizebox{0.5\textwidth}{!}{\includegraphics{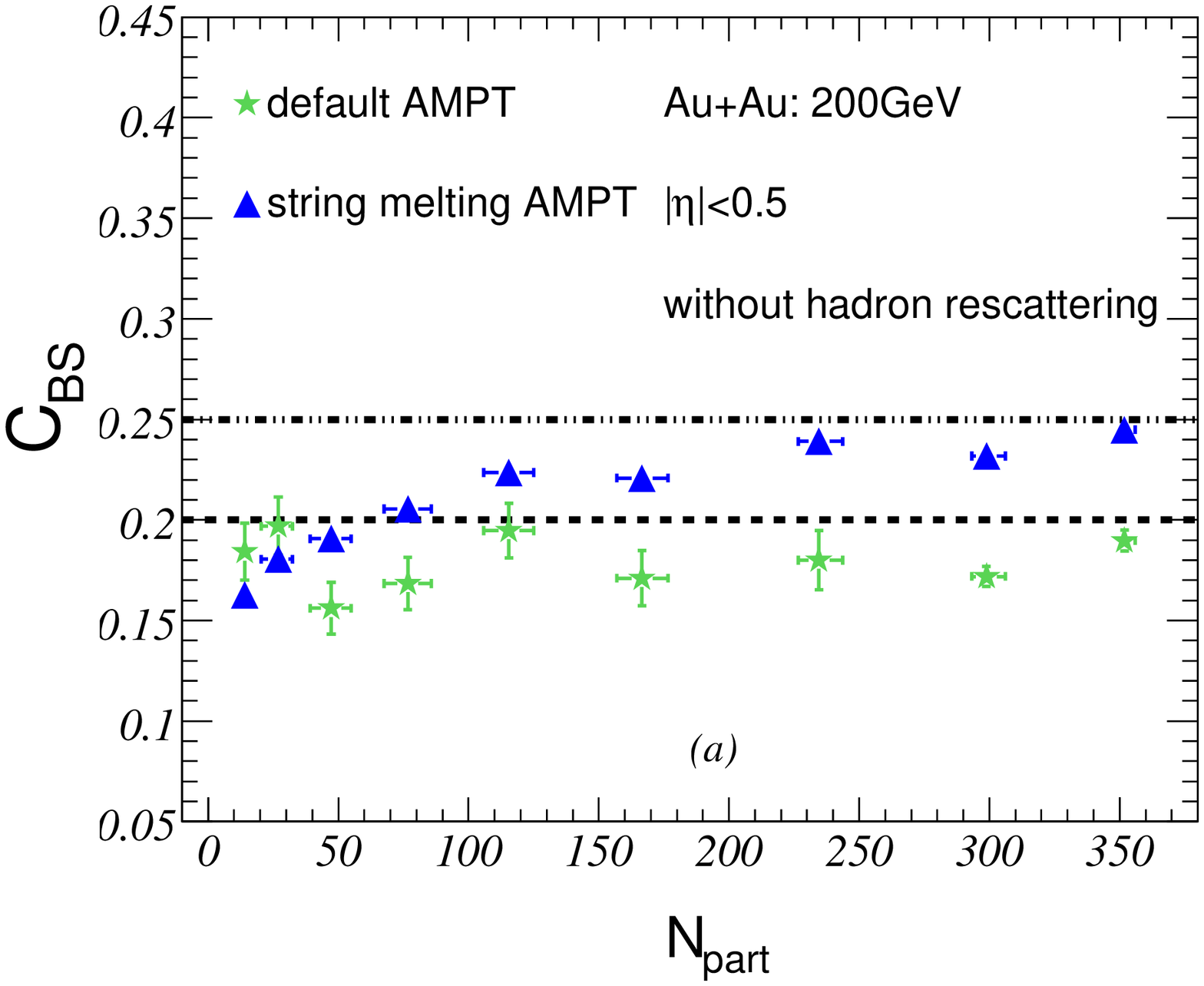}}
\resizebox{0.5\textwidth}{!}{\includegraphics{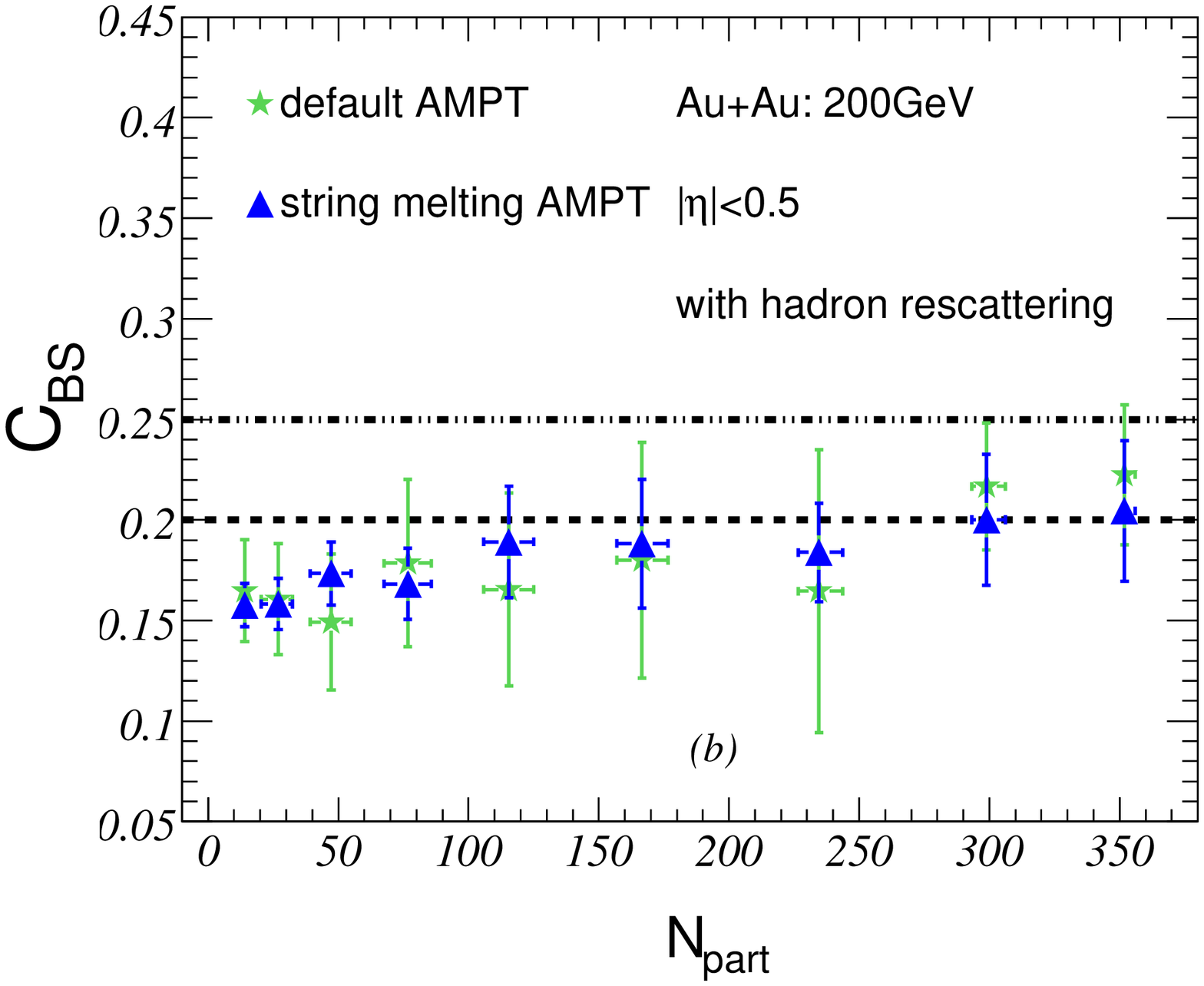}}
\caption[]{ $C_{BS}$ of Kaons and Protons combination as a
function of $N_{part}$ at the $\eta_{max}$ = 0.5 in the default
AMPT and the string melting AMPT without(a) or with hadronic
rescattering(b). } \label{fig5}
\end{figure}

Finally, we try to choose a particle subset to see how much
hadronic rescattering effect on the $C_{BS}$. Here, the subset
only includes Kaons and Protons. We find the value of $C_{BS}$
goes down to 0.2 from Fig.~\ref{fig5}. But after hadronic
rescattering we get the similar results that hadronic rescattering
finally obliterates the signals of partonic matter completely.

\section{Summary}\label{sum}

We have studied the dependence of the $C_{BS}$ as a function of
the $N_{part}$ with a multi-phase transport model. At $\eta_{max}$
= 0.5, we find the hadronazation makes the $C_{BS}$ value drop,
but we still obtain the residual signal. However, after the
hadronic rescattering, the residual signal will be washed out. In
order to analyze the effect of hadronic rescattering, we choose a
special particle group which only includes kaons and protons, but
the result does not help. Up till now, there are no powerful
fluctuation probes of the deconfined matter in the experiment,
perhaps both hadronization and hadronic rescattering effects may
be responsible for the disappearance of the signals.

Authors appreciate the organizers of SQM2007. This work was
supported in part by the National Natural Science Foundation of
China under Grant No. 10610285 and 10705044, the Knowledge
Innovation Project of the Chinese Academy of Sciences under Grant
No. KJCX2-YW-A14 and and KJXC3-SYW-N2,  the Shanghai Development
Foundation for Science and Technology under Grant No. 05XD14021.
And we thank Information Center of Shanghai Institute of Applied
Physics of Chinese Academy of Sciences for using PC-farm.


\section*{References}

\begin{thebibliography}{34}

\bibitem{Karsch_NPA698}
  F.~Karsch,
   {\it Nucl.\ Phys.\  A} {\bf 698}, 199c (2002).

\bibitem{BRAHMS_NPA757}
  I.~Arsene, {\it et al.}, [BRAHMS Collaboration],
   {\it Nucl.\ Phys.\  A} {\bf 757}, 1 (2005);
  B.~B.~Back, {\it et al.}, [PHOBOS Collaboration],
  {\it Nucl.\ Phys.\  A} {\bf 757}, 28 (2005);
  J.~Adames, {\it et al.}, [STAR Collaboration],
  {\it Nucl.\ Phys.\  A} {\bf 757}, 102 (2005);
  S.~S.~Adler, {\it et al.}, [PHENIX Collaboration],
  {\it Nucl.\ Phys.\  A} {\bf 757}, 184 (2005).

\bibitem{Stodolsky_PRL75}
  L.~Stodolsky,
  {\it Phys.\ Rev.\ Lett.}  {\bf 75}, 1044 (1995).

\bibitem{Shuryak_PLB423}
  E.~V.~Shuryak,
  {\it Phys.\ Lett.\ B}  {\bf 423}, 9 (1998).
  [arXiv:help-ph/9704456].

\bibitem{Bleicher_NPA638}
  M.~Bleicher, {\it et al.},
  {\it Nucl.\ Phys.\  A} {\bf 638}, 391 (1998).

\bibitem{Koch_PRL95}
  V.~Koch, A.~Majumder, and J.~Randrup
  {\it Phys.\ Rev.\ Lett.}  {\bf 95}, 182301 (2005).

\bibitem{Haussler_PRC73}
  S. Haussler, H. St\"ocker, and M. Bleicher
  {\it Phys.\ Rev.\ C}  {\bf 73}, 021901(R) (2006).

\bibitem{BZhang_PRC72}
  Z.~W.~Lin, C.~M.~Ko, B.~A.~Li and S.~Pal
  {\it Phys.\ Rev.\ C}  {\bf 72}, 064901 (2005).

\bibitem{XNWang_PRD44}
  X.~N.~Wang, and M.~Gyulassy
  {\it Phys.\ Rev.\ D}  {\bf 44}, 3501 (1991).

\bibitem{BZhang_CPC109}
  B.~Zhang
  {\it Comput.\ Phys.\ Commun}  {\bf 109}, 193 (1998).

\bibitem{BZhang_PRC61}
  B.~Zhang, C.~M.~Ko,{\it et al.},
  {\it Phys.\ Rev.\ C}  {\bf 61}, 067901 (2000).

\bibitem{BAnderson_PR97}
  B.~Andersson, G.~Gustafson,{\it et al.},
  {\it Phys.\ Rep}  {\bf 97}, 31 (1983).

\bibitem{ZWLin_PRC65}
  Z.~W.~Lin, C.~M.~Ko,
  {\it Phys.\ Rev.\ C}  {\bf 65}, 034904 (2002);
  Z.~W.~Lin, C.~M.~Ko,{\it et al.},
  {\it Phys.\ Rev.\ Lett.}  {\bf 89}, 152301 (2002).

\bibitem{BALi_PRC52}
  B.~A.~Li, C.~M.~Ko,{\it et al.},
  {\it Phys.\ Rev.\ C}  {\bf 52}, 2037 (1995).

\bibitem{GLMa_PLB647}
  G.~L.~Ma, Y.~G.~Ma, S.~Zhang {\it et al.},
  {\it Phys.\ Lett.\ B}  {\bf 647}, 122 (2007);
  G.~L.~Ma, S.~Zhang, Y.~G.~Ma {\it et al.}, {\it Phys. Lett.B} {\bf 641}, 362 (2006).

\bibitem{JHChen_PRC74}
  J.~H.~Chen, Y.~G.~Ma, G.~L.~Ma {\it et al.},
  {\it Phys.\ Rev.\ C}  {\bf 74}, 064902 (2006).

\bibitem{SZhang_PRC76}
  S.~Zhang, G.~L.~Ma, Y.~G.~Ma {\it et al.},
  {\it Phys.\ Rev.\ C}  {\bf 76}, 014904 (2007).

\bibitem{Majumder_0}
  A.~Majumder, V.~Koch and J.~Randrup,
  arXiv: nucl-th/0510037

\end{thebibliography}
\end{document}